\documentstyle[prd,preprint,aps,epsfig,floats]{revtex}

\def\D0{D\O~}
\def\thisday{\today}

\def\beq{\begin{equation}}
\def\eeq{\end{equation}}
\def\bdm{\begin{displaymath}}
\def\edm{\end{displaymath}}
\def\bea{\begin{eqnarray}}
\def\eea{\end{eqnarray}}

\def\url#1{\mbox{\href{#1}{\sf #1}}}

\def \etv{E_T\!\!\!\!\!\!/~~} 
\def \snuL{\tilde \nu_{\tau_{\small L}}} 
\def \snuR{\tilde \nu_{\tau_{\small R}}} 
\def \eebar{e^+e^-}
\def \snu{\tilde \nu}
\def \snuF{\tilde \nu_{\tau_1}} 
\def \snuS{\tilde \nu_{\tau_2}} 

\begin{document}
\draft
\tighten


\title{New Production Mechanism of Neutral Higgs Bosons with \\
Right scalar tau neutrino as the LSP
}

\author{C.-L. Chou, ${}^1$ H.-L. Lai  ${}^2$ and C.-P. Yuan ${}^3
$\thanks{
On leave of absence 
from Department of Physics and Astronomy, Michigan state University, 
Michigan 48824, U.S.A.}} 
\address{Institute of Physics, Academia Sinica, Taipei 11529, Taiwan
 ${}^1$\\ 
Department of Physics, National Cheng-Kung University, Tainan 70101, 
Taiwan ${}^2$ \\ 
Theory Division, CERN, CH-1211, Geneva, Switzerland ${}^3$} 

\date{\thisday}

\maketitle
\thispagestyle{empty}

\begin{abstract}

Motived by the neutrino oscillation data, we consider the lightest tau 
sneutrino $\tilde \nu_{\tau_1}$ (which is mostly the right tau sneutrino) 
to be the lightest supersymmetric particle (LSP) in the framework of the 
minimal supersymmetric Standard Model. Both the 
standard and the non-standard trilinear scalar coupling terms are included 
for the right tau sneutrino interactions. The decay branching ratio of 
$\tilde \nu_{\tau_2} \rightarrow \tilde \nu_{\tau_1}+ h^0$ can become so 
large that the production rate of the lightest neutral Higgs boson ($h^0$) 
can be largely enhanced at electron or hadron colliders, either from the 
direct production of $\tilde \nu_{\tau_2}$ or from the decay of charginos, 
neutralinos, sleptons, and the cascade decay of squarks and gluinos, etc.
Furthermore, because of the small LSP annihilation rate, 
$\tilde \nu_{\tau_1}$ can be a good candidate for cold dark matter.  

\pacs{PACS number(s): 
12.60.Jv,  
14.80.Cp,  
14.80.Ly,  
14.60.St \\
{}\\
{} \hfill CERN-TH/2000-182}

\end{abstract}


\setcounter{footnote}{0}
\renewcommand{\thefootnote}{\arabic{footnote}}


\section{Introduction}

Supersymmetry (SUSY) has been largely studied as a 
possible framework for the theory beyond the Standard Model (SM).  
It provides a natural solution to the hierarchy problem, the generation
of the electroweak symmetry breaking, as well as the grand unification of  
the gauge coupling constants \cite{mssm}.
Among the various supersymmetric models, the minimal supersymmetric 
Standard Model (MSSM) is the one studied most extensively in the 
literature \cite{mssm}.  
In addition to the SM particles, it consists of the 
supersymmetric partners of the SM particles, called sparticles.
All renormalizable interactions,
including both SUSY conserving and (soft) SUSY breaking terms, 
are assumed to conserve the $B-L$ global symmetry, which then 
results in the conservation of the $R$-parity, $R=(-1)^{3(B-L)+2S}$. 
($B$ denotes the baryon number, $L$ the lepton number, and $S$ the spin
of the (s)particle.)
 Accordingly, all the SM particles have 
 even $R$-parity and the sparticles have odd $R$-parity.  
 This fact has an important consequence, namely, if the initial particles 
 of a scattering
process are the SM particles, then sparticles can only be produced 
 in pairs.  $R$-parity conservation also implies the existence of a
  stable sparticle, called the lightest supersymmetric particle (LSP).  
  The LSP is absolutely stable and cannot decay.   
 
Recently, the Super-K neutrino experiment presented the high precision 
neutrino oscillation data which strongly suggests the existence of the 
neutrino mass \cite{SuperK}.
Many SUSY models have been proposed to account for a
reasonable set of neutrino masses 
with bi-maximal mixing among the three family neutrinos \cite{king}.
The low energy effective theory of these SUSY models can be summarized
in the supersymmetric extension of the see-saw model, in which 
a right-handed neutrino superfield is added for each family in the 
framework of the MSSM \cite{haber}.
As proposed in \cite{king}, the observed bi-maximal mixing in neutrino
data can be explained by the single right-handed neutrino dominance
mechanism, which assumes that the light effective Majorana matrix come
predominantly from a single right-handed neutrino which generates some
particular textures of the Yukawa couplings in the superpotential of the 
low energy effective theory.
With that example in mind, we assume the bi-maximal mixing among the 
three family neutrinos does not necessarily imply a large mixing among 
different flavor sneutrinos because the attendant R-parity conserving 
soft-supersymmetry breaking terms are not fixed by the generation of
neutrino masses or mixings. Furthermore, the trilinear scalar coupling
terms introduced by supersymmetry breaking can be large in some SUSY 
models.
For simplicity, we shall only consider a one family model, 
though its phenomenology
is expected to be applicable to three family models after properly
including possible mixing factors.

Despited that the conventional supersymmetric see-saw models predict 
heavy right sneutrinos, we presume the existence of SUSY models,
in which the interactions of right sneutrinos 
with left sneutrinos at the weak scale are
described by Eqs.(\ref{eqn:sneutrinoMixing}),
(\ref{eqn:mixingangle}) and (\ref{eqn:ACcoupling})
with R-parity conservation in the framework of MSSM,
which may or may not include lepton number violation interactions.
We study the scenario that
 the lightest sneutrino, mostly the right tau sneutrino
 {$\snuR$}, is the LSP.
 To further simplify our discussion, we also assume that the mixings 
 among the three generation 
 sneutrinos are small enough that the dominant effect to
 collider phenomenology comes from the interaction of left and right
tau sneutrinos. 
In section II, we give our assumptions and formalism for the $\snuF$-LSP
 scenario.
 In section III, we discuss the possible large production rate of the
 lightest neutral Higgs boson predicted by this model
 at electron and hadron colliders.
 We also show that with a small left-right mixing,
 a $\snuR$-like $\snuF$ can be a good candidate for cold dark matter.   
 In section IV, we consider in details some $\eebar$ collider
 phenomenology for the $\snuF$-LSP model. 
 Section V contains our conclusion.  

\section{$\snuF$-LSP scenario} 

In our one family model,
the scalar tau neutrinos, like scalar quarks and 
scalar leptons, can mix and form mass eigenstates, and  
\begin{eqnarray} 
\left( \begin{array}{c} 
           \snuS \\ 
           \snuF 
         \end{array} \right) 
           &=&\left( 
   	\begin{array}{cc} 
                \cos\theta_{\snu} &  \sin\theta_{\snu} \\ 
               -\sin\theta_{\snu}  & \cos\theta_{\snu}       
	\end{array} 
   \right) 
   \left( \begin{array}{c} 
         \snuL \\ 
         \snuR^* 
         \end{array} 
   \right)   
, \label{eqn:sneutrinoMixing} 
\end{eqnarray} 
with 
\begin{equation} 
\tan 2\theta_{\snu}= {2 \Delta m^2 \over m^2_{\snuL} - m^2_{\snuR}} \, ,
\label{eqn:mixingangle}
\end{equation} 
 where $\snuL$ and $\snuR$ stand for the left and right tau
 sneutrinos, and  $m^2_{\snuL}$ and $m^2_{\snuR}$ are the corresponding 
 soft masses, respectively. 
 We consider the case that the parameter $\Delta m^2$ mainly comes
from the soft 
 SUSY breaking effect originated from the trilinear scalar couplings in
 \begin{equation} 
\Delta {\cal L} = A_{\mbox{v}} H_2 {\tilde L_3} \snuR + 
C_{\mbox{v}} H_1^* {\tilde L_3} \snuR+ \mbox{c.c.}  
\label{eqn:ACcoupling} 
\end{equation} 
where $A_{\mbox{v}} (C_{\mbox{v}})$ is the standard
 (non-standard) trilinear scalar coupling, $H_1$ and $H_2$ are 
 the two Higgs doublets and $\tilde L_3$ is the third generation scalar 
 lepton doublet. (To simplify our discussion, 
 we have assumed that the contribution to 
$\Delta m^2$ from the $\mu$-term in the superpotential is much smaller than 
that from $\Delta {\cal L}$, because of the small Yukawa coupling
 of tau neutrino.)
The parameter $\Delta m^2$ arises after 
 the Higgs doublets $H_1$ and $H_2$ acquiring their vacuum expectation
  values $v_1$ and $v_2$, and 
\begin{eqnarray} 
\Delta m^2& \approx & v \left( A_{\mbox{v}} \sin\beta - 
C_{\mbox{v}}\cos\beta \right) \, , 
\label{eqn:deltam2}
\end{eqnarray} 
with $v=\sqrt{v_1^2+v_2^2} \approx 176 \mbox{ GeV}$
 and the angle $\beta = \tan^{-1}{({v_2 /v_1})}$.
As shown in Eq.(\ref{eqn:ACcoupling}), we have included the non-standard
trilinear scalar coupling $C_{\mbox{v}}$ in the
soft-supersymmetry-breaking scalar potential to extend the applicability
of our effective model. As to be shown later, a special value of 
$C_{\mbox{v}}$ can lead to distinct collider signatures.
 The masses of the mass eigenstates $\snuF$ and $\snuS$ are
 given by
$m_{\snuF,\snuS}^2=({m^2_{\snuL}+m^2_{\snuR} \mp
\sqrt{(m^2_{\snuL}-m^2_{\snuR})^2+4(\Delta m^2)^2}})/2 $,
with $m_{\snuF} < m_{\snuS}$. 
In terms of the sneutrino masses,
the sneutrino mixing factor $\sin2\theta_{\tilde \nu}$ can 
be written as: 
\begin{equation}
\sin2\theta_{\tilde \nu} = {2\Delta m^2 \over m^2_{\snuS}-m^2_{\snuF}}
\, .
\label{eqn:mixing2}
\end{equation}
Since we require a positive 
mass for the lightest tau sneutrino $\snuF$, we must have 
the mass parameters satisfy the following constraint:
\begin{equation}
m^2_{\snuL}m^2_{\snuR}>(\Delta m^2)^2 \, .
\label{eqn:massconstraint}
\end{equation}

\begin{figure}[h]
\vspace*{-1.3cm}
\begin{center} 
\leavevmode{\epsfxsize=4.00truein\epsfbox{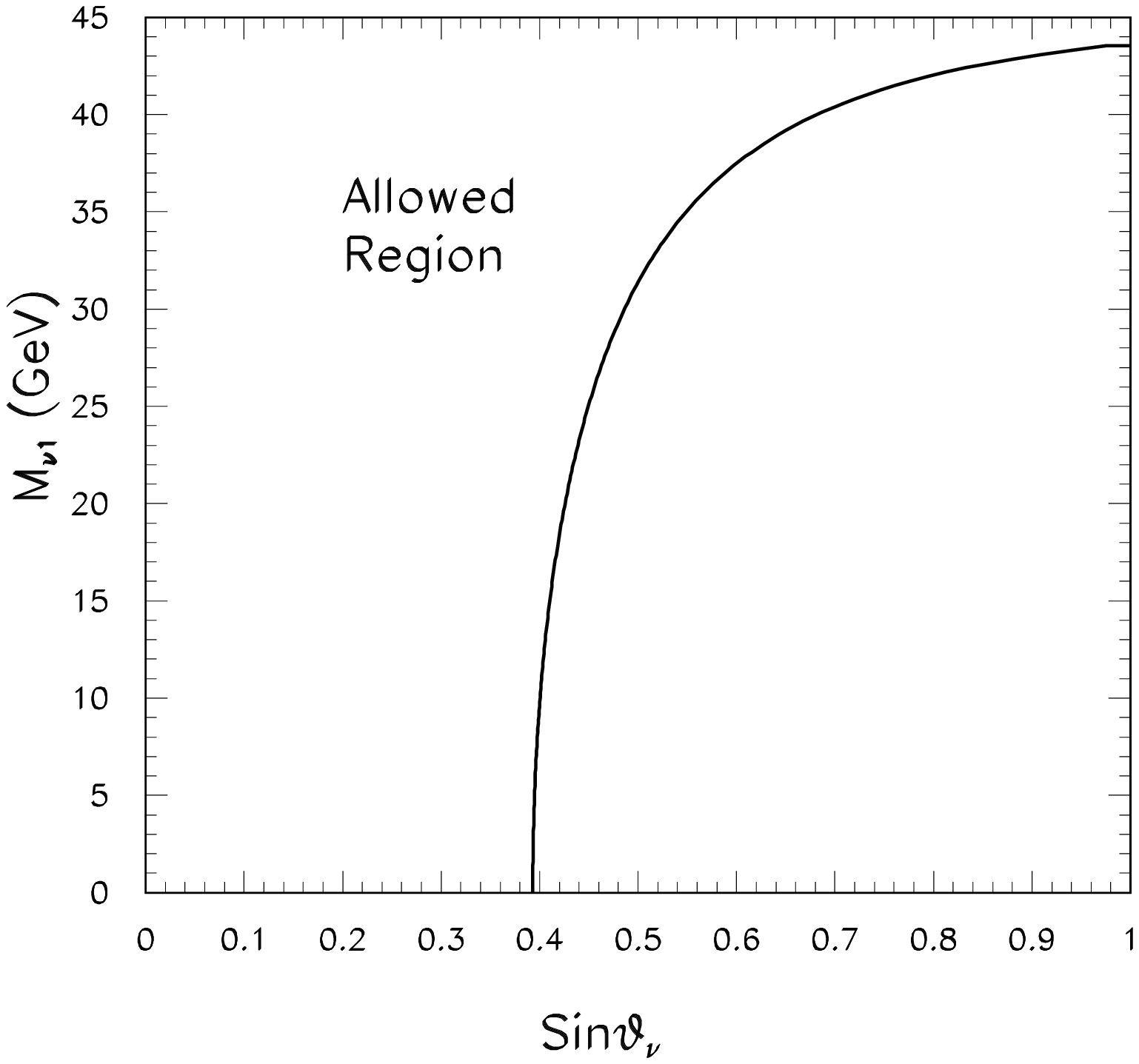}} 
\end{center} 
\caption[fig:Aconst]{
The allowed parameter space (the left-hand side of the curve) on 
the $(\sin\theta_{\snu}, m_{\snuF})$ plane, for $m_{\snuF}<m_Z/2$. 
}
\label{fig:Aconst}
\end{figure}

Under the $\snuF$-LSP scenario, the LSP can be generated either through
 the direct productions in high energy collision or through the decay 
 of heavy sparticles, 
 such as the next-to-lightest supersymmetric particle (NLSP)
  and the heavy tau sneutrinos $\snuS$. 
It can also be 
 pair-produced through the $Z$ boson decay if the $\snuF$ mass is smaller 
 than one half of the $Z$ mass.  The partial decay width
  for the $Z$ boson decaying into a $\snuF \snuF^*$ pair is given at 
  the tree level as: 
 \begin{equation} 
\Gamma( Z \rightarrow \snuF \snuF^*) = {{ \alpha_{em}(\cot\theta_W + 
\tan\theta_W)^2 }\sin^4\theta_{\snu}\over 48 }m_Z\left[1-{4m^2_{\snuF} 
\over m_Z^2}\right]^{3/2} \, ,
\label{Z2RR}
\end{equation} 
where $\theta_W$ is the weak mixing angle and $\alpha_{em}$ is the fine
structure constant evaluated at the $Z$-mass scale $m_Z$.
Clearly, the partial decay width 
$\Gamma( Z \rightarrow \snuF \snuF^*)$
 depends on the $\snuF$ mass and the mixing angle $\sin\theta_{\snu}$.
From the experimental data at LEP and SLC, the invisible decay channel
of $Z$ boson is bounded from above, and
$
\Delta \Gamma(Z \rightarrow \mbox{invisible}) < 2 \mbox{ MeV} 
$ {}\cite{Zdata}.
Though this experimental constraint is automatically satisfied for 
the lightest tau sneutrino mass
$m_{\snuF}$ to be larger than one half of the $Z$-boson mass, 
it does not excludes the possibility 
that $m_{\snuF}$ can be very small as compared to $m_Z/2$.
Indeed, we find that $m_{\snuF}$ can take any small value 
as long as the mixing angle 
$\sin\theta_{\snu}$ is smaller than about $0.39$.
 Because ${\snuL}$ carries the electroweak quantum number and
 ${\snuR}$ is a singlet field, a SUSY model usually predicts the soft mass  
 parameter $m_{\snuL}$ to be larger than $m_{\snuR}$ at the weak scale,
 after including the running effect of the renormalization 
 group equations. This case is assumed hereafter.
From Eqs.(\ref{eqn:mixingangle}) and (\ref{eqn:massconstraint}),
we conclude that the soft mass parameter
$\Delta m$ should be less than 
$m_{\snuL} \, \sqrt{\tan\theta_{\tilde \nu}}$, 
and $m_{\snuL} < m_{\snuR } \, {\cot\theta_{\tilde \nu}}$.
Consequently, as an example, for ${m_{\snuL}}$ to be 
 200 GeV, the upper bound 
 on ${\Delta m}$ is about $130 \mbox{ GeV}$, which
  imposes a constraint on the values of 
  $A_{\mbox{v}}$ and $C_{\mbox{v}}$ for
  a fixed $\tan \beta$.
(Following the method in Ref. \cite{ufb},
we have checked that in this model, there is no useful bound on 
the value of $A_{\mbox{v}}$, assuming $C_{\mbox{v}}=0$, from 
requiring the absence of dangerous charge and color
breaking minima or unbounded from below directions.)
  For $\sin\theta_{\snu} > 0.39$, a larger $\sin\theta_{\snu}$ requires 
a larger $m_{\snuF}$, e.g.,
when  $\sin\theta_{\snu}$ is 0.5, the minimal allowed value for $m_{\snuF}$
 is $ 32 \mbox{ GeV}$.  Our results are 
 summarized in Fig.\ref{fig:Aconst}, which shows the allowed 
 parameter space on the $(\sin\theta_{\snu}, m_{\snuF})$ 
 parameter plane. 
 (Only the allowed range for $m_{\snuF} < m_Z/2$ is
 shown. For $m_{\snuF} > m_Z/2$, $\sin\theta_{\snu}$ can take any value 
 within 1.)

\section{New Production Mechanism of Higgs Bosons at Colliders}

The trilinear scalar couplings for the neutral Higgs bosons and 
tau sneutrinos can be 
derived from Eqs.(\ref{eqn:sneutrinoMixing}) and
(\ref{eqn:ACcoupling}) as:
\begin{eqnarray} 
\Delta {\cal L} &=& \{H^0(A_{\mbox{v}}\sin\alpha - C_{\mbox{v}}\cos\alpha) 
+ h^0(A_{\mbox{v}}\cos\alpha+C_{\mbox{v}}\sin\alpha)\} 
   \{\sin2\theta_{\snu}(\snuS\snuS^* - \snuF\snuF^*) \nonumber \\
 &+& \cos2\theta_{\snu}(\snuF\snuS^*+\snuS\snuF^*)\}/{\sqrt 2} 
 + iA^0(A_{\mbox{v}}\cos\beta+C_{\mbox{v}}\sin\beta)
(\snuS\snuF^* - \snuF\snuS^*)/{\sqrt 2} \, ,  
\label{eqn:vvhcoupling} 
\end{eqnarray} 
where $h^0$ is the lightest CP-even neutral Higgs boson, 
$H^0$ denotes the other CP-even neutral Higgs boson and $A^0$ is the CP-odd
 neutral Higgs particle. The phase angle $\alpha$ defines the mass 
eigenstates of $h^0$ and $H^0$.

When the mixing angle $\theta_{\snu}$ is small,
 the lightest sneutrino $\snuF$ (which is almost $\snuR$-like)
 predominantly 
interacts with $\snuS$ (which is almost $\snuL$-like) 
via the scalar interactions, cf.
$\Delta {\cal L}$.
Hence, to study the production of the LSP $\snuF$,  
it is desirable to first examine
 the decay and the production of $\snuS$. 

\begin{figure}[t] 
\begin{center} 
\leavevmode{\epsfxsize=4.00truein\epsfbox{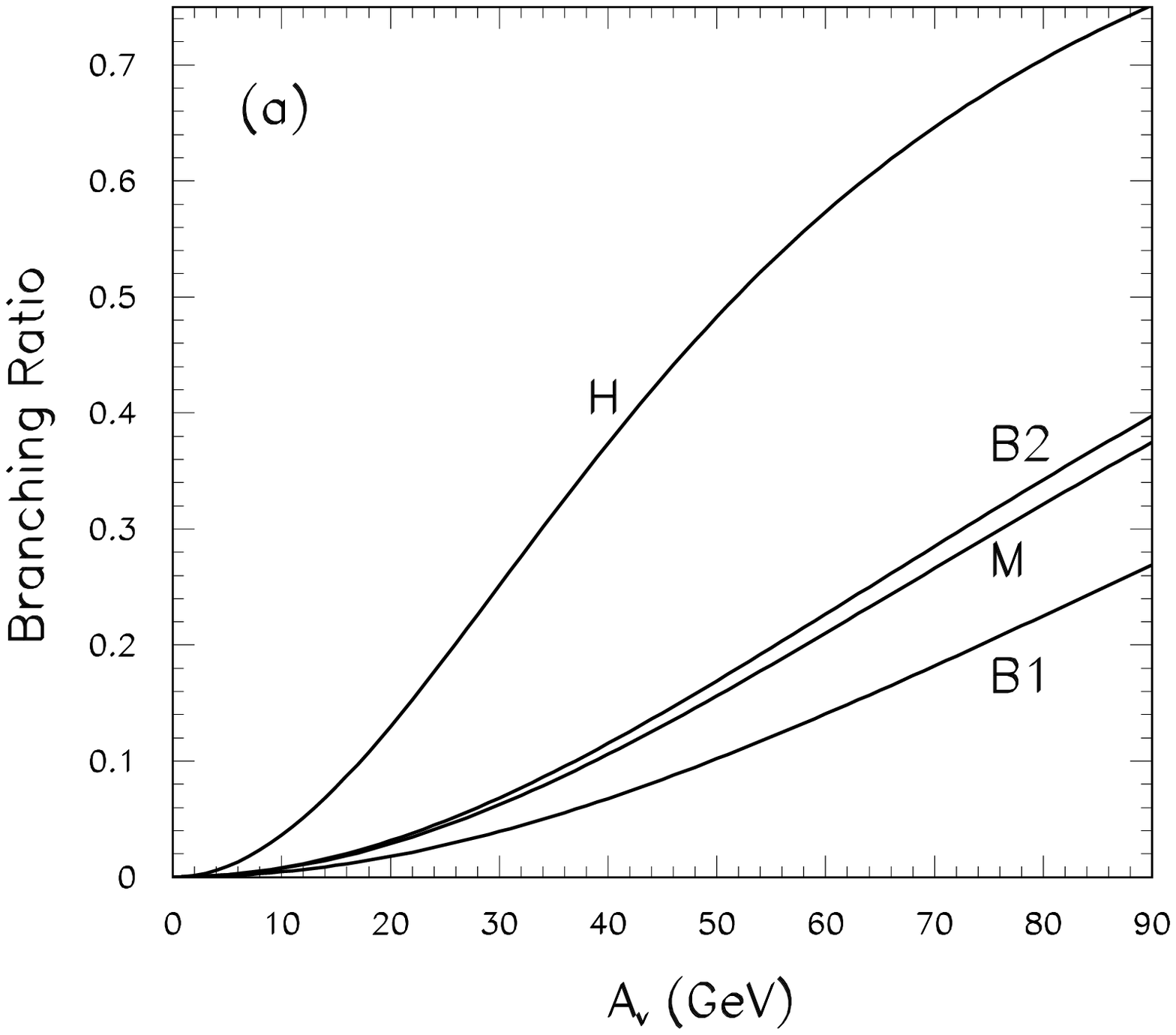}} 
\end{center} 
\begin{center} 
\leavevmode{\epsfxsize=4.00truein\epsfbox{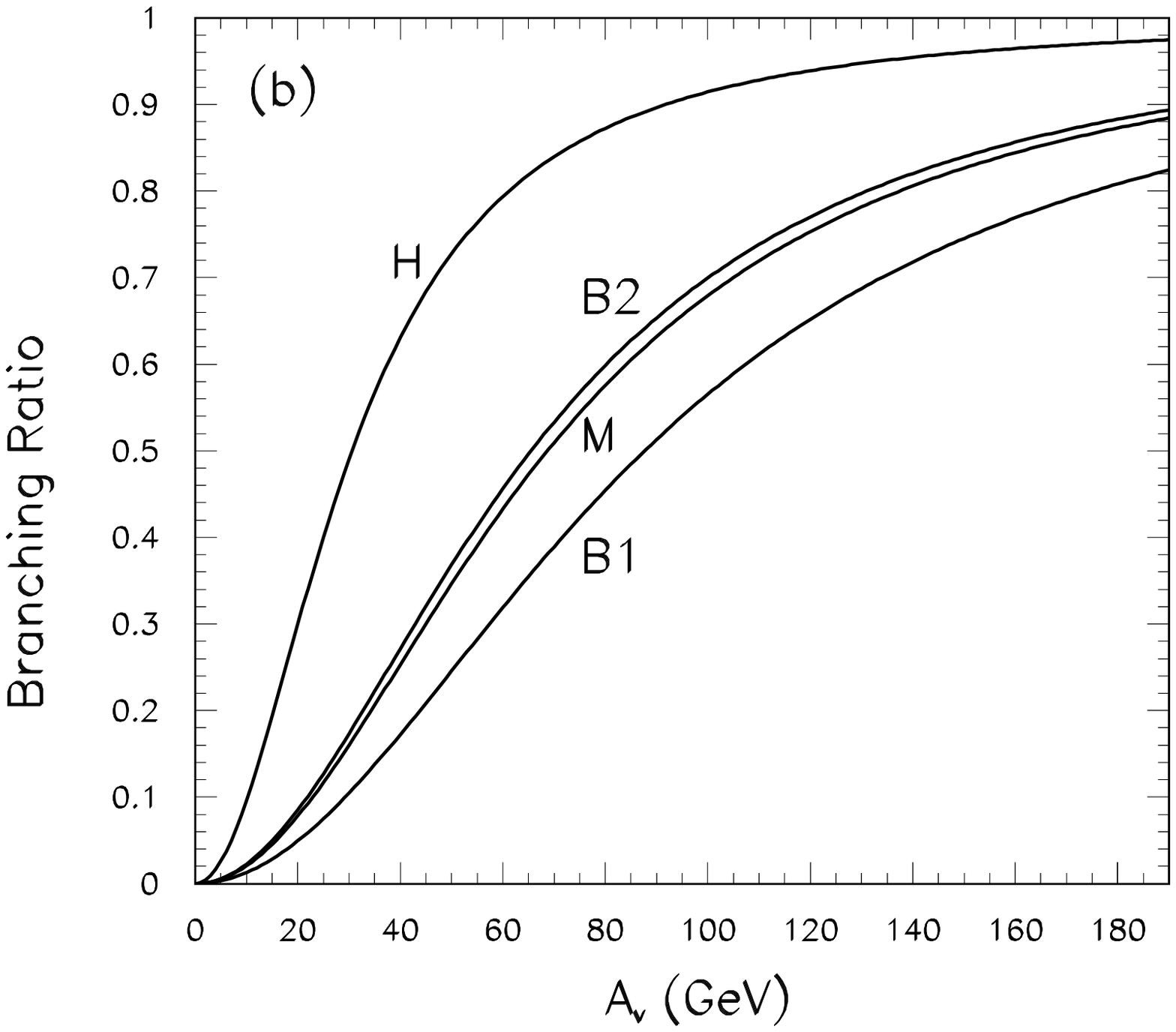}} 
\end{center} 
\caption{The decay branching ratio BR($\snuS \rightarrow 
\snuF + h^0$), as a function of $A_{\mbox{v}}$, 
with (a) $C_{\mbox{v}}=0$, $\tan\beta=2$, and (b)
$C_{\mbox{v}} = A_{\mbox{v}} \tan\beta$, $\tan\beta=2$.
A detailed explanation is given in the text.} 
\label{fig:SLnuDCY} 
\end{figure}

When $\snuS$ is produced at colliders, it may 
decay into the tau neutrino $\nu_{\tau}$ and the 
neutralino $\tilde\chi^0_1$ (or $\tilde\chi^0_2$),
 or into the Higgs particle $h^0$ 
and the lightest sneutrino $\snuF$, assuming the other modes 
either are forbidden by mass relation or have negligible partial decay
widths.)
  The branching ratios (BR)
of these two decay modes depend on the SUSY parameters. 
For illustration, we give their tree level 
partial decay widths as follows: 
 \begin{eqnarray} 
\Gamma(\snuS \rightarrow \snuF h^0) &=& {\cos^22\theta_{\snu} 
|A_{\mbox{v}} \cos\alpha + 
C_{\mbox{v}} \sin\alpha|^2 \over 32 \pi m_{\snuS} } \nonumber \\
&\times& \sqrt{ \left[ \left(1+{m_{h^0} \over m_{\snuS}}\right)^2- 
{m_{\snuF}^2 \over m^2_{\snuS} }  \right]  
\left[\left(1-{m_{h^0} \over m_{\snuS}} \right)^2 - 
{ m_{\snuF}^2 \over m^2_{\snuS} } \right] }  \, ,
\label{eqn:decay2h}\\   
\Gamma( \snuS \rightarrow {\nu_{\tau}}  \tilde\chi^0_j) &=& {\alpha_{em} 
m_{\snuS} \cos^2\theta_{\snu}\over 8} 
\big|{V_{1j}\over \cos\theta_W}-{V_{2j} \over \sin\theta_W}\big|^2 
\left(1-{m^2_{\tilde\chi_j^0} \over m^2_{\snuS}}\right)^2, 
\label{eqn:decay2chi0} 
\end{eqnarray} 
where $V_{1j}$ and $V_{2j}$, for $j=1,2,3,4$, 
 denote the matrix elements 
of the diagonalizing matrix for the neutralino mass matrix \cite{kane}.
Based on Eqs.(\ref{eqn:decay2h}) and (\ref{eqn:decay2chi0}), we plot
in Fig.\ref{fig:SLnuDCY}(a)
  the branching ratio BR($\snuS \rightarrow \snuF + h^0$) 
for $m_{\snuS}=200$ GeV, $m_{\snuF}=20$ GeV and $m_{h^0}=130$ GeV,
as a function of $A_{\mbox{v}}$, with  $C_{\mbox{v}} = 0$
and $\tan \beta =2$. 
(The constraint from the invisible decay width 
of $Z$ boson requires $A_{\mbox{v}} < 96 \mbox{ GeV}$, 
with $\sin\theta_{\snu} < 0.42$,
cf. Fig.\ref{fig:Aconst} and Eq.(\ref{eqn:mixing2}).) 
Four curves for different 
neutralino mixing scenarios are plotted. The first two 
curves (B1 and B2) are for $\tilde \chi^0_1$-NLSP
to be Bino-like, in which we have assumed 
the common soft SUSY breaking masses 
$m_1=\mbox{100 GeV}$, 
   $m_2=\mbox{200 GeV}$ and $\tan\beta=2$, but with different 
   $\mu$ values: $\mu = 500, -500$\,GeV. 
The third curve (M) is 
   for the mixed-type $\tilde \chi^0_1$-NLSP scenario, in which 
    $\mu=-100$\, GeV.
The last one (H) is for the 
Higgsino-like $\tilde \chi^0_1$-NLSP 
    scenario, in which $m_1 = 200\mbox{ GeV}$, $m_2 = 400\mbox{ GeV}$ 
    and $\mu = -100 \mbox{ GeV}$. As shown in the figure, the branching
     ratio increases as $A_{\mbox{v}}$ increases. 
Furthermore, due to the small Yukawa 
     coupling of the $\snuS$-$\mbox{higgsino}$-$\nu_{\tau}$ interaction, 
     the branching ratio BR($\snuS \rightarrow \snuF + h^0$) increases 
rapidly  under the Higgsino-like $\tilde \chi^0_1$-NLSP scenario.  

The amusing feature of the model is that the lightest 
neutral Higgs boson $h^0$ can be largely
produced at either electron or hadron colliders from the decay of 
$\snuS$, which is mainly $\snuL$ when the left-right mixing angle is
small. Since a left
sneutrino $\snuL$ carries electroweak quantum number, it
can be produced directly in collisions, or indirectly from
the decay of charginos, neutralinos, sleptons, and the cascade decay of
squarks and gluinos, etc. There are plenty of studies in the
literature to show that the production rates of 
the above mentioned sparticles at the current and future colliders 
can be very large, depending on the SUSY parameters \cite{tevlhc}. 
In that case, our model would predict a large production rate of 
events including either single $h^0$ or multiple $h^0$'s, provided that
BR($\snuS \rightarrow \snuF + h^0$) is large enough.
Many of the events including $h^0$'s can 
also contain single or multiple isolated leptons and/or photons
with large transverse momentum, 
so it will not be difficult to trigger on such events experimentally. 
Due to the limited space in this short Letter, we cannot explore all the
interesting possibilities in details for various colliders. Instead, we
shall illustrate the above observation for the future Linear Collider
(LC) in the next section.

Before closing this section, we remark that,
in contrast to the $\snuL$-LSP scenario,
a $\snuR$-like $\snuF$ can be a good candidate for the cold dark 
matter (CDM).
The left sneutrino had been suggested in the literature
to be the LSP in the MSSM \cite{LsneutrinoLSP}. They can annihilate
rapidly in the early universe via s-channel $Z$-boson 
and t-channel neutralino and chargino exchanges. To reduce the LSP 
annihilation and obtain an acceptable relic abundance, 
it was proposed that the left sneutrinos should be either as light 
as $m_{\tilde\nu} \approx 2\mbox{ GeV}$ or as heavy as 
$550 \mbox{ GeV} < m_{\tilde\nu} < 2300 \mbox{ GeV}$ \cite{LsnMass}.  
However, both of these proposals have been excluded by 
experiments. The light $\tilde \nu_L$ scenario was
excluded by the measurements of $Z$ decay width,
and the heavy $\tilde \nu_L$ scenario was excluded by the 
Heidelberg-Moscow direct detection experiment 
\cite{ExByRecoil}.  
Since a $\snuR$-like $\snuF$  interacts with other 
particles mainly through the left-right sneutrino mixing 
or the trilinear
 scalar coupling $\snuR$-$\snuL$-$h^0$, the LSP annihilation cross sections 
 are generically small due to the presence of the small 
$\sin^4\theta_{\snu}$ or $\sin^2\theta_{\snu}$ factors
coming from the mixing effect or the couplings.
Comparing to the $\tilde\nu_{L}$-LSP annihilation in the 
ordinary MSSM \cite{cdm}, 
the $\snuF$-LSP annihilation rate 
via the exchange of $Z$-boson, neutralinos, or charginos, 
is suppressed by 
a factor of $\sin^4\theta_{\snu}$ because the mixing of 
${\snuR}$ and ${\snuL}$ yields a factor of $\sin\theta_{\snu}$
in the scattering amplitude.
In addition to the usual MSSM processes, ${\snuF}$ can also annihilate
via an s-channel Higgs boson to produce light fermion pairs, 
whose scattering amplitude is suppressed by a factor of
$\sin\theta_{\snu}$.
Notice that the above
 rate can strongly depend on $\tan \beta$ because of the coupling of 
Higgs boson and fermions (such as bottom quarks).
In lack of a complete SUSY model which gives the mass spectrum of the
sparticles, and the ${\snuF}$ annihilation rate depends on the details of
the MSSM parameters, 
we only remark that with the additional suppression factor  
discussed above, a ${\snuR}$-like LSP ${\snuF}$ can be a good candidate
for CDM.

\section{ $\snuF$-LSP phenomenology at the LC} 

In this section, we consider a simple example to illustrate
the interesting phenomenology of our model expected at high energy 
colliders.

The tree-level cross section for the production of $\snuS$ pair
 in $e^+ e^-$ collision is
\begin{eqnarray} 
\sigma(e^+ e^- \rightarrow \snuS \snuS^*)=
{\pi \alpha_{em}^2 \cos^2\theta_{\snu}\over 6S}
{1-4\sin^2\theta_W+8\sin^4\theta_W \over \sin^42\theta_W}
\left[1-{4m^2_{\snuS} 
\over S}\right]^{3/2}\left[1-{m_Z^2\over S}\right]^{-2} \, , 
\label{eqn:ee2LL} 
\end{eqnarray} 
through the s-channel $Z$-exchange diagram.
($\sqrt{S}$ is the center-of-mass energy of the collider.) 
As a comparison, the tree-level cross sections for the direct 
productions of $\snuF$ 
 in $e^+ e^-$ collision associated with ${\snuS^*}$ or ${\snuF^*}$
are 
 \begin{eqnarray} 
\sigma(e^+ e^- \rightarrow \snuF \snuS^*)  &=& 
\sigma(e^+ e^- \rightarrow \snuF^* \snuS) \nonumber \\
&=&{\pi \alpha_{em}^2 \sin^22\theta_{\snu}\over 24 S}
{1-4\sin^2\theta_W+8\sin^4\theta_W \over 
\sin^42\theta_W} \nonumber \\ 
&\times& \left[ \left( 1-{m^2_{\snuS}-m^2_{\snuF} \over S}\right) ^2 - 
{4m^2_{\snuF} \over S}\right]^{3/2}  \left[1-{m_Z^2\over S}\right]^{-2}   
 \, , \label{eqn:ee2RL} \\ 
\sigma (e^+ e^-   \rightarrow {\snuF} {\snuF^*}) &=& 
{\pi \alpha_{em}^2 \sin^4\theta_{\snu} 
\over 6 S}{1-4\sin^2\theta_W+8\sin^4\theta_W \over \sin^42\theta_W} 
\nonumber \\ 
&\times& \left[ 1-{4m^2_{\snuF} \over S}\right]^{3/2} 
\left[1-{m_Z^2\over S}\right]^{-2} \, .
\label{eqn:ee2RR} 
\end{eqnarray} 
Hence, the direct productions of ${\snuF}$ could be highly 
suppressed when the factor $\sin^4\theta_{\snu}$ is much smaller than 1. 
Apart from a different phase space factor,
the direct production of ${\snuF} {\snuF^*}$ 
is smaller than the production of ${\snuS} {\snuS^*}$ 
by a factor of $\sin^4\theta_{\snu}/\cos^2\theta_{\snu}$.

As mentioned in the previous section,  $\snuS$ could decay 
into $\snuF h^0$ or $\nu_{\tau} \tilde \chi^0_j$.   
With a large BR($\snuS \rightarrow \snuF h^0$), it may be possible to  
observe the $2h^0 + \etv$ signal originated from the production of  
the $\snuS$ pair.
To test this scenario, we need to calculate
the SM rate for the process
 $e^+ e^- \rightarrow h^0  h^0  \nu_i {\bar \nu_i}$, where
 $\nu_i$ (${\bar \nu_i}$) is the left-handed
  neutrino (anti-neutrino) for the $i$th family. 
Since the final 
  state neutrinos, like the LSPs, carry away energy, the above SM process
is the intrinsic background to the detection of the signal event 
$e^+ e^- \rightarrow \snuS \snuS^* \rightarrow h^0 h^0 \snuF \snuF^*$,
because both processes produce the event signature of 
 $e^+ e^- \rightarrow 2 h^0 +\etv$.
Due to the large suppression factor from the
4-body phase space, the cross section for
 the SM process is typically small. For example, when the Higgs mass
  is 130 GeV, the SM rate is about $0.03 \, \mbox{fb}$,
for $\sqrt{S}=500 \, \mbox{GeV}$.
  On the other hand, for a 200 GeV $\snuS$, with small left-right tau 
sneutrino mixing effect (i.e. $\sin\theta_{\snu} \sim 0$), the 
tree-level cross section for the $\snuS$-pair production is about
12 fb.  This relatively large production rate can lead to an enhancement 
 in the Higgs boson pair signal,
provided BR($\snuS \rightarrow \snuF h^0$) is large enough,
cf. Fig.\ref{fig:SLnuDCY}.
 
It is also interesting to consider a special case with 
$C_{\mbox{v}}=A_{\mbox{v}} \tan\beta $,
so that $\Delta m^2$ vanishes and $\snuL$ does not
 mix with $\snuR$. Hence, the direct production of 
 $\snuF$ (now a pure $\snuR$) vanishes and  
 $\snuF$ is predominantly produced through 
the decay of $\snuS \rightarrow \snuF h^0$. 
As shown in Fig.\ref{fig:SLnuDCY}(b), 
the branching ratio BR($\snuS \rightarrow \snuF + h^0$) is 
generally higher than that in 
Fig.\ref{fig:SLnuDCY}(a) because of the inclusion of the 
non-standard soft SUSY breaking parameter $C_{\mbox{v}}$. 
For instance, 
assuming the same mass parameters as the Higgsino-like 
$\tilde \chi^0_1$-NLSP scenario given above,
the BR($\snuS \rightarrow \snuF + h^0$) is 
about 0.1 for $A_{\mbox{v}}=10\mbox{ GeV}$, and 0.9 for 
$A_{\mbox{v}} = 90\mbox{ GeV}$ with $\tan \beta=2$, which 
results in the production cross section of
 $e^+ e^- \rightarrow \snuS \snuS^* \rightarrow h^0h^0\snuF \snuF^*$
to be about 0.12 fb and 11 fb, respectively.
When comparing to the SM rate, the signal rate of our model can be
 larger by about a factor of 300.
With a larger $\tan \beta$, the branching ratio of 
$\snuS \rightarrow \snuF + h^0$ increases and more $2 h^0+\etv$
signal events are expected.

Thus far, we have only shown that the $\snuF$-LSP signal rate of  
 $2h^0 +\etv$ production can be much larger than the SM rate.
 However, to distinguish our $\snuF$-LSP scenario
from the ordinary MSSM scenario with $\tilde \chi^0$-LSP, we also need to 
compare our signal rate with that predicted by the ordinary MSSM.
When the decay mode 
$\tilde\chi^0_2 \rightarrow \tilde\chi^0_1 + h^0$ is available,
 the MSSM $2h^0 +\etv$ signature mainly comes from the pair production 
 of $\chi^0_2$ in $e^+ e^-$ collisions.
The event rate of 
$e^+ e^- \rightarrow \tilde\chi^0_2 \tilde\chi^0_2
\rightarrow h^0 h^0 \tilde\chi^0_1 \tilde\chi^0_1$
depends on the detail of the MSSM parameters. For example, assuming the 
usual scenario that $\tilde\chi^0_1$ is almost Bino-like, the partial
decay width of 
$\tilde\chi^0_2 \rightarrow h^0 \tilde\chi^0_1 $
will come from one-loop corrections.
Since the difference in the masses of $\chi^0_2$ and $\chi^0_1$
is large enough (greater than $m_{h^0}$) for
the other tree-level decay modes of $\chi^0_2$ to be opened, the
branching ratio 
BR($\tilde\chi^0_2 \rightarrow h^0 \tilde\chi^0_1$) 
is generally small.
Furthermore, if $\tilde\chi^0_2$ is Higgsino-like, the cross section of 
$e^+ e^- \rightarrow \tilde\chi^0_2 \tilde\chi^0_2$ 
at a 500 GeV collider is expected to be
small, at the order of fb or smaller, for MSSM mass parameters 
to be at the order of a few hundreds of GeV. For a gaugino-like 
$\tilde\chi^0_2$, the above cross section can increase by a factor of 10.
A detailed comparison between our model and the ordinary MSSM for the
event rate of $e^+e^- \rightarrow 2 h^0 \etv$ is beyond the scope of this
short Letter.

\section{conclusion} 

Motivated by the neutrino oscillation data, we 
study a low energy effective theory
in which the interactions of right sneutrinos 
with left sneutrinos at the weak scale are
described by Eqs.(\ref{eqn:sneutrinoMixing}),
(\ref{eqn:mixingangle}) and (\ref{eqn:ACcoupling})
with R-parity conservation in the framework of MSSM.
For simplicity, we assume the mixings among the three generation 
 sneutrinos are small enough that the dominant effect to
 collider phenomenology comes from the interaction of the 
left and right tau sneutrinos, which
 mix via the soft SUSY breaking effect arising 
from the trilinear scalar couplings of scalar Higgs doublets and sleptons.  
We find that the mass of the lightest tau sneutrino
$\snuF$ can take any value (even smaller than $m_Z /2$)
 to agree with the $Z$ decay width measurement, provided that
the sneutrino mixing parameter $\sin\theta_{\snu}$ is smaller than 
 $0.39$. In that case, $\snuF$ is almost the right tau sneutrino $\snuR$,
 and becomes a good candidate to be the LSP of the model as well as 
 the cold dark matter.
Because $\snuR$ mainly interacts with $\snuL$ through Higgs boson, 
the branching ratio BR($\snuS \rightarrow \snuF + h^0$) can be large.
It results in a large production rate of single $h^0$ or multiple
$h^0$'s in electron or hadron collisions via the decay of $\snuS$,
which is $\snuL$-like and can be copiously produced from the decay of 
charginos, neutralinos, sleptons, and the cascade decay of
squarks and gluinos, etc.
Hence, the events including Higgs boson(s) can also contain single or
multiple isolated leptons and/or photons
with large transverse momentum, which can make it easy to
trigger on such events experimentally. 
Given the possible large production rate of the above mentioned
sparticles, the production rate of the lightest neutral Higgs boson is
expected to be largely enhanced from its SM rate.
For example, with the trilinear couplings
$C_{\mbox{v}}=A_{\mbox{v}}\tan\beta$, 
the left and right sneutrino do not mix,
and the BR($\snuS \rightarrow \snuF + h^0$) is approaching 
to 1 as $A_{\mbox{v}}$ increases. In that case, 
the signal rate ($\sim 12$ fb) of
$e^+e^- \rightarrow 2h^0+\etv$ at a 500 GeV LC
is enhanced by a factor of 400, as compared to its SM rate.
 
\section*{Acknowledgments}

CPY thanks L. Diaz-Cruz and Y. Okada for discussions, and 
G.L. Kane for useful suggestions and a critical reading of the manuscript.
We are also grateful to the warm hospitality of the Center for Theoretical 
Science in Taiwan where part of this work was completed.
This work is in part supported by the National Science Council 
in Taiwan and the National Science Foundation in the USA
under the grant PHY-9802564. 

\vspace*{3.5mm}
\noindent
\underline{Note Added:}
\linebreak\hspace*{2.5mm}

After posting this manuscript to the xxx-archives, stamped as 
hep-ph/0006313, we noticed that in a new paper, hep-ph/0006312,
several supersymmetry breaking mechanisms were proposed for generating 
light sneutrinos. CPY thanks N. Weiner for explaining the SUSY models 
proposed in Ref. \cite{neal}, and pointing out an error in the previous
version of the manuscript.




\newpage

\end{document}